\documentclass[12pt,a4paper]{article}
\usepackage{a4}
\usepackage{cite}
\usepackage{graphicx}
\usepackage{amssymb}
\usepackage{amsmath}
\usepackage{epsfig}
\usepackage[tmargin=1cm,bmargin=1cm,lmargin=2cm,rmargin=2cm]{geometry}

\newcommand{\qb}{\ensuremath{\overline q}}
\newcommand{\qp}{\ensuremath{q \prime}}
\newcommand{\qbp}{\ensuremath{ {\overline q} \prime }}
\newcommand{\rarrow}{\ensuremath{\rightarrow}}
\newcommand{\qqpZqqp}{\ensuremath{q \qp \rarrow q \qp}}
\newcommand{\qqZqq}{\ensuremath{q q \rarrow q q}}
\newcommand{\qqbZqqb}{\ensuremath{q \qb \rarrow q \qb}}
\newcommand{\qqbZqpqbp}{\ensuremath{q \qb \rarrow \qp \qbp}}
\newcommand{\qqbZgg}{\ensuremath{q \qb \rarrow gg}}
\newcommand{\ggZqqb}{\ensuremath{gg \rarrow q \qb}}
\newcommand{\qgZqg}{\ensuremath{qg \rarrow q g}}
\newcommand{\ggZgg}{\ensuremath{gg \rarrow g g}}
\nopagebreak

\begin{document}
\pagestyle{empty}
\baselineskip 22pt

\begin{flushright}
SINP/TNP/2009/14\\
{\tt arXiv:0903.0202[hep-ph]}
\end{flushright}
\vskip 65pt
\begin{center}
{\large \bf
Di-jet production at the LHC through unparticles
} \\
\vspace{8mm}
{\bf
Neelima Agarwal$^{a}$
\footnote{neel1dph@gmail.com},
M. C. Kumar$^{b,c}$
\footnote{mc.kumar@saha.ac.in},
Prakash Mathews$^b$
\footnote{prakash.mathews@saha.ac.in},
\\
V. Ravindran$^d$
\footnote{ravindra@hri.res.in},
Anurag Tripathi$^d$
\footnote{anurag@hri.res.in}
}\\
\end{center}
\vspace{10pt}
\begin{flushleft}
{\it
a) Department of Physics, University of Allahabad, Allahabad 211002, India. \\
b) Saha Institute of Nuclear Physics, 1/AF Bidhan Nagar,
Kolkata 700 064, India.\\
c) School of Physics, University of Hyderabad, Hyderabad 500 046, India.\\
d) Regional Centre for Accelerator-based Particle Physics,\\ Harish-Chandra Research Institute,
 Chhatnag Road, Jhunsi, Allahabad 211 019, India.\\
}
\end{flushleft}
\vspace{10pt}
\centerline{\bf Abstract}

We report the phenomenological impact of unparticles in the production of 
di-jet at the LHC.  We compute the scalar, spin-1 and spin-2 unparticle 
contributions to the dijet cross sections and present our results in 
different kinematical distributions.  We find that the scalar unparticle 
contribution is dominant over that of the spin-1 and spin-2 unparticles 
for the same coupling values.

\vskip12pt
\vfill
\clearpage

\setcounter{page}{1}
\pagestyle{plain}

The Large Hadron Collider (LHC), will explore the origin of the 
spontaneous symmetry breaking which is responsible for giving 
masses to gauge bosons in the Standard Model (SM). Its main effort 
will be on the discovery of the elusive Higgs boson which is a 
signature of the spontaneous symmetry breaking and is the last
missing particle of the SM.  Even if Higgs Boson is discovered, 
many questions remain unanswered in the SM.  This indicates to 
the existence of some New Physics (NP) at high energies beyond 
the SM.  As the LHC will operate at high energies never attained 
before in any experiment,  signals of NP beyond the SM could be 
discovered here. There are many possible NP scenarios, such as, 
supersymmetry, extra-dimensions etc. There 
are other interesting possibilities as well that
have been thought of, especially involving scale or conformal
symmetries. "Conformal collider physics" was recently discussed in
\cite{Hofman:2008ar}. Similarly the newly envisaged
unparticle model \cite{Georgi:2007ek}, though does not address the 
problems of the SM, is interesting in its 
own right. This model is based on the curious possibility of
scale invariant degrees of freedom coupling to the SM fields
at low energies. In this model, at very high energies
the SM couples weakly to a hidden sector, called Banks-Zaks (BZ) 
sector \cite{Banks:1981nn}, via exchange of 
heavy particles of mass M.
Effective theory below M contains interaction terms of the
form
\begin{equation}
\frac{1}{M^k} {\cal O}_{SM} {\cal O}_{BZ}\,,
\end{equation}
where $ {\cal O}_{SM}$ and $ {\cal O}_{BZ}$ are operators constructed
out of SM and BZ fields respectively.
 This hidden sector is proposed to have a
nontrivial infrared fixed point.  One can take an effective 
field theory approach and integrate out high energy
degrees of freedom.  As the modes are integrated out the 
renormalization group flow takes us close to the IR fixed
point. Near the fixed point, scale invariance
 emerges in the 
hidden sector below a scale $\Lambda_u$.
The above interaction term below $ \Lambda_u$ matches to 
\begin{equation}
\frac{ {\Lambda_u}^{d_{BZ}-d_u } } {M^k}   { \cal O}_{SM} {\cal O}_u.
\end{equation}
The operators $ {\cal O}_u $ are scale invariant and  
scale with momenta with some scaling dimension $d_u$, which
depends on the operator. 
$d_{BZ}$ is the mass dimension of $ {\cal O}_{BZ}$ operator. 
Apart from an overall normalization, the scale invariance fixes the 
two-point function of unparticle operators without requiring 
any detailed knowledge of the theory at high energies. 
Unparticle operators can have different tensor structures, such as
scalar, vector or tensor. 
Here, in this study, we will consider scalar, spin-1 and spin-2 unparticles. 
The effective interaction for scalar and spin-1 unparticle consistent with 
the SM gauge symmetries are:
\begin{equation}
\label{Lint}
{\cal L}_{int} \supset   \frac{ \lambda_{{s_1}}}{4 \Lambda_u^{{d_s}}}~
 F^a_{\mu \nu} F^{a \mu \nu}~ {\cal O}_u
+ \frac{\lambda_{ {s_2}}}{\Lambda_u^{{d_s-1}}}~{\bar \psi} \psi ~{\cal O}_u\, ,
\end{equation}
\begin{equation}
\label{Lintv}
{\cal L}_{int} \supset
 \frac{\lambda_{ {v}}}{\Lambda_u^{{d_v-1}}}~{\bar \psi} \gamma_\mu \psi ~{\cal O}_u^\mu\, ,
\end{equation}
where $\lambda_i$ are the dimensionless coupling constants of the 
scalar ($i=s_{1,2}$) and vector ($i=v$) unparticles, $d_i$ are the 
scaling dimension of scalar ($i=s$) and vector ($i=v$) unparticle
operators.
 
For the spin-2 unparticle, we assume that the SM fields couple to 
unparticle operator $ O^{\mu \nu}_u $  via the SM energy momentum 
tensor $T_{\mu \nu}$ :
\begin{eqnarray}
\label{Lintt}
{\cal L}_{int} \supset {\lambda_t \over \Lambda^{d_t}_{u}} T_{\mu \nu} ~ 
O^{\mu \nu}_u\,,
\end{eqnarray}
Tensor operator $O^{\mu \nu}_{u}$ which is traceless 
and symmetric  has a scaling dimension $d_t$.  Unitarity imposes constraint 
$d_s > 1$ on the scaling dimension of scalar unparticle \cite{Mack:1975je} 
and scale invariance restricts $d_v,d_t \geq 3$ \cite{Grinstein:2008qk}. 
Scalar, tensor and vector unparticle propagators  are given respectively 
by \cite{Georgi:2007si, Cheung:2007zza,Grinstein:2008qk} 
\begin{eqnarray}
\label{props}
\int d^4x ~ e^{-i k\cdot x}
\langle 0 |
     T O_u(x) O_u(0)
|0 \rangle 
  &=&
    -i~
       C_S \frac {  \Gamma(2 -d_s)} {4^{d_s-1} 
       \Gamma(d_s)}
       ( - k^2)^{d_s-2} 
\\[2ex]
\label{propv}
\int d^4x ~ e^{-i k\cdot x}
\langle 0 |
     T O^{\mu \nu}_u(x) O^{\alpha \beta}_u(0)
|0 \rangle &=&-i~
       C_T \frac {  \Gamma(2 -d_t)} {4^{d_t-1} 
       \Gamma(d_t+2)}
       ( - k^2)^{d_t-2}  
\nonumber\\[2ex]
  &&  \times  \left[ d_t(d_t-1)
        (g_{\mu \alpha} g_{\nu \beta} + \mu \leftrightarrow \nu)
      + \ldots \right]\,,
\\[2ex]
\label{propt}
\int d^4x ~ e^{-i k\cdot x}
\langle 0 |
     T O_u^\mu (x) O_u^\nu (0)
|0 \rangle &=&+i~
       C_V \frac { (d_v-1)  \Gamma(2 -d_v)} {4^{d_v-1}
       \Gamma(d_v+1)}
       ( - k^2)^{d_v-2}
\nonumber\\[2ex]
  &&  \times  \left[ g^{\mu \nu} - \frac{2(d_v-1)}{d_v-1} \frac{k^\mu k^\nu}{k^2} \right]\,,
\end{eqnarray}
where $C_S,C_T,C_V$ are overall normalisation constants.  The terms given 
by ellipses in Eq.\ (\ref{propv}) and the terms proportional to $k^\mu 
k^\nu $ in Eq.\ (\ref{propt}), do not contribute to di-jet production.

Recently unparticle phenomenology, in the context of present and 
future colliders has been explored in great detail \cite{Cheung:2008xu}. 
At the LHC, in the case of unparticles, the study of di-jet was 
considered in \cite{Alan:2007rg} by considering only the $g g \rightarrow 
g g$ sub process assuming that the gluon flux would help this channel to 
dominate over the others.   
This analysis of \cite{Alan:2007rg} was prior to the results of 
\cite{Grinstein:2008qk}, which in the context of unparticles 
pointed out the lower bounds on operator dimensions and dictates 
the tensor structure of the unparticle propagators.  
In our analysis we have taken note of the above points and consider 
all the sub processes that would contribute to the di-jet via the 
exchange of spin-0, spin-1 and spin-2 unparticles. Including all the subprocess
is important as the gluon and quark operators couple to the scalar 
unparticle operators with different powers of the unparticle scale 
$\Lambda_u$ (see Eq.\ (\ref{Lint}-\ref{Lintt})).  Hence the enhancement
due to gluon flux at the LHC could be suppressed by the addition
power of $\Lambda_u$ in the gluonic coupling to scalar unparticles.

The LHC can provide a testing ground for physics of unparticles if 
$ \Lambda_u$ is of the order of a TeV. There are
various important channels available at the LHC
to explore the new physics at TeV scales,  namely production of di-leptons, 
isolated photon pairs, di-jets etc.  The importance of di-lepton and
di-photon production channels in the context of unparticles were
already studied in detail in \cite{Mathews:2007hr} and 
\cite{Kumar:2007af,Kumar:2008dn} respectively. 
In this article we will study the effects of 
scalar and spin-2 unparticles on the di-jet production rates
at the LHC if the scale $ \Lambda_u$ is of the order of a TeV.
Di-jet production is an important discovery mode and many 
studies in the context of various new physics scenarios have 
been carried out, namely, SUSY searches    
\cite{Randall:2008rw}, searches of the low mass strings 
\cite{Anchordoqui:2008di}.
Di-jet production has been used to probe spin-2 Kaluza-Klein gravitons
appearing in the extra dimensional models 
\cite{Mathews:1999iw,Atwood:1999qd,Ghosh:1999ex}.
 
To lowest order in strong coupling constant, di-jets arise from 
$2 \rightarrow 2 $ scattering of partons.  In the unparticle model
the dijets could be produced due to the exchange of unparticles
between SM particles.  The partons in the final state hadronize to
give two jets in the detectors. Signals of NP can be discovered 
because of the deviations they produce over the SM contributions.
The unparticles can contribute through
intermediate states as well as via real emission.
The former one can interfere with the SM contributions, while the
later can lead to missing energy in the final state. 
In this article we will restrict ourselves to the effects
coming from spin-0, spin-1 and spin-2 intermediate states. 
The parton 
level $2 \rightarrow 2$ subprocesses in the SM and in the unparticle 
scenario
are
\begin{align*}
q \qp \rarrow q \qp  &\hskip 2pt&  q q \rarrow  q q   &\hskip 2pt&  q \qb \rarrow q \qb \\
q \qb \rarrow \qp \qbp  && q \qb \rarrow g g  &&   gg \rarrow q \qb   \\ 
qg \rarrow qg   && gg \rarrow gg  
\end{align*}
Here we have used primes to distinguish quark flavours.  For the 
scalar case, not all of the above processes contribute.  The first 
term in Eq.\ (\ref {Lint}) describes the coupling of gauge fields 
to scalar unparticles giving $gg \rightarrow gg $ process. The second 
term couples fermions to scalar unparticles, which  allows subprocesses 
that have only fermions in the initial and the final states.  We shall 
study the effects of these two terms in Eq.\ (\ref{Lint}) separately.
In the case of vector unparticles, there are no processes involving 
gluons (Eq.\ (\ref{Lintv})) that would contribute to the dijet process.

The leading order SM matrix elements are of order $g_s^2$, where $g_s$ 
is the strong coupling constant and those in the  unparticle model are 
of order $ \kappa^2 $, where $ \kappa = \lambda_t / \Lambda_u^{d_t}$, 
~ $\lambda_{s_1}/ \Lambda_u^{d_s}$, $\lambda_{s_2}/ \Lambda_u^{d_s-1},
\lambda_v/\Lambda_u^{d_v-1}$.
The matrix element square takes the following form:
\begin{equation*}
g_s^4 \left | {\cal M}_{SM}  \right |^2 
~+~ \kappa^4        \left | {\cal M}_{u}  \right |^2
~+~ g_s^2 \kappa^2 ( {\cal M}_{SM}    {\cal M}_u^* +  {\cal M}_{SM}^*
    {\cal M}_u )   
\end{equation*} 
%
where the interference of the SM with the unparticle
mediated processes will be sensitive to phase coming from $(-k^2)^{d}$ in the
propagators given in Eq.\ (\ref{props}-\ref{propt}).
In the table (\ref{smm}), we list the matrix elements 
square, summed (averaged) over final (initial) state colors
and spins, for SU(N) gauge theory with fermions in the fundamental 
representation. The SM ones agree 
with those existing in the literature 
\cite{Combridge:1977dm}.  We have not listed the subprocesses
such as $\qb~\qb \rightarrow \qb~\qb$, $\qb~ \qbp \rightarrow \qb~ \qbp$ and
$\qb g \rightarrow \qb g$ as they can be obtained
from the rest using charge conjugation.  
In addition, there are processes that are 
related by crossing symmetry {\em viz.} $\qb~ \qp \rightarrow
\qb~ \qp$ is obtained from $q~ \qp \rightarrow q~ \qp$.

The matrix elements for pure unparticle contribution and interference 
with SM  for spin-0 coupling through the first term in (\ref{Lint}) 
are given below (see Eq.\ (\ref{upintf})).  Here, only $ gg \rightarrow gg$ 
subprocess contributes.  The factor $ (-k^2)^{d-2} $ in propagators
is complex for a $s$-channel propagator and is real for 
$u$- and $t$-channel propagators. 
\begin{eqnarray}
|{\cal M}_{u}|^2 &\stackrel{gg \rarrow gg}{=}&  
\frac{1}{16(N^2-1)} ~ \Big( {\cal D}_u Re({\cal D}_s) ~ s^2~u^2 
+ {\cal D}_t Re({\cal D}_s)~ s^2~t^2 
\nonumber\\[2ex]
&&+ {\cal D}_t {\cal D}_u ~t^2~ u^2 \Big) 
+\frac{1}{16} \left( {\cal D}_u^2 ~  u^4 
+ {\cal D}_t^2 ~  t^4 
+ |  {\cal D}_s|^2 ~  s^4  \right) \\[2ex] 
2 {\cal R}e \Big({\cal M}_{SM} {\cal M}^{*}_{u}\Big) 
&{{\stackrel{gg \rarrow gg}{=}}}& 
 -~\frac{N}{2(N^2-1)} ~ \Big(  Re({\cal D}_s) ~ \frac{s^4}{ut}  
                             + {\cal D}_u ~ \frac{u^4}{st}  
                             + {\cal D}_t ~ \frac{t^4}{us} \Big) 
\label{upintf}
\end{eqnarray}
where
\begin{equation}
\label{sDs}
{\cal D}_s  =
    - C_S \frac {  \Gamma(2 -d_s)} {4^{d_s-1} 
       \Gamma(d_s)}
       ( - s)^{d_s-2} .
\end{equation}
Table (\ref{smatrix}) contains the corresponding matrix element square for 
spin-0 unparticle interacting via the  second term in Eq.\ (\ref{Lint}).
In table (\ref{tt}), we give the matrix element square for
spin-2 unparticles with $ {\cal D}_s$ as defined by 
\begin{equation}
\label{tDs}
{\cal D}_s  =
    - C_T \frac {  \Gamma(2 -d_t)} {4^{d_t-1} 
       \Gamma(d_t+2)}
       ( - s)^{d_t-2}  
   d_t(d_t-1).
\end{equation}
The matrix elements for spin-1 are given in table (\ref{somatrix}) with 
corresponding ${\cal D}_s$ defined by
\begin{equation}
\label{vDs}
{\cal D}_s  =
    + C_V \frac {(d_v-1) \Gamma(2 -d_v)} {4^{d_v-1}
       \Gamma(d_v+1)}
       ( - s)^{d_v-2}
\end{equation}
The $t$ ($u$)-channel propagators can be obtained by the replacement 
$s \rightarrow t(u)$. 

We now study invariant mass $Q$ distribution, namely $ d \sigma /d Q $, 
of the di-jet for the LHC with a center of mass of energy $ \sqrt{S} = 
14 ~$ TeV.  We have implemented all the parton level matrix element 
squares in a Monte Carlo based code that can accommodate all the 
experimental cuts relevant for the phenomenological study.  A $1/2$ 
factor for sub processes involving identical particles in the final
state has been taken care of, when the angular integration is done 
over the range $-1 \le \cos(\theta) \le 1$.  We use the leading order 
(LO) CTEQ 6L parton distribution functions (PDF) with the corresponding 
value of LO strong coupling constant $\alpha_s(M_Z) = 0.118$ and $5$ 
light quark flavours. The factorization scale $ \mu_F$ that appears 
in the PDFs and the renormalization scale $ \mu_R$ in  $\alpha_s 
(\mu_R ^2) $ are identified with a single scale $Q$.  In accordance 
with the CMS  \cite{Bhatti:2008hz}, we restrict the jets to satisfy 
rapidity cut $ |\eta| < 1$ and the transverse momentum cut $p_{T} > 50$ 
GeV for each final state jet.   We choose the scale $ \Lambda_u $
to be $ 2~$ TeV below which the scale invariance in the BZ sector sets in.
The dimensionless coupling constants are chosen to be 
$ \lambda_t, \lambda_{s1}, \lambda_{s2}, \lambda_v   = 0.9$ 
(see Eq.\ (\ref{Lint}-\ref{Lintt}))
for all our phenomenology and we fix $ C_T,C_S,C_V =1$ 
appearing in the normalization of the propagators given in 
Eq.\ (\ref{props}-\ref{propt}).
Notice that the di-jet production at the LHC has a huge SM back ground
unlike Drell-Yan or the di-photon production case.
We will also study the sensitivity of our results to the scaling dimensions 
of the unparticle operators.

\begin{figure}[htb]
\vspace{1mm}
\centerline{
\epsfig{file=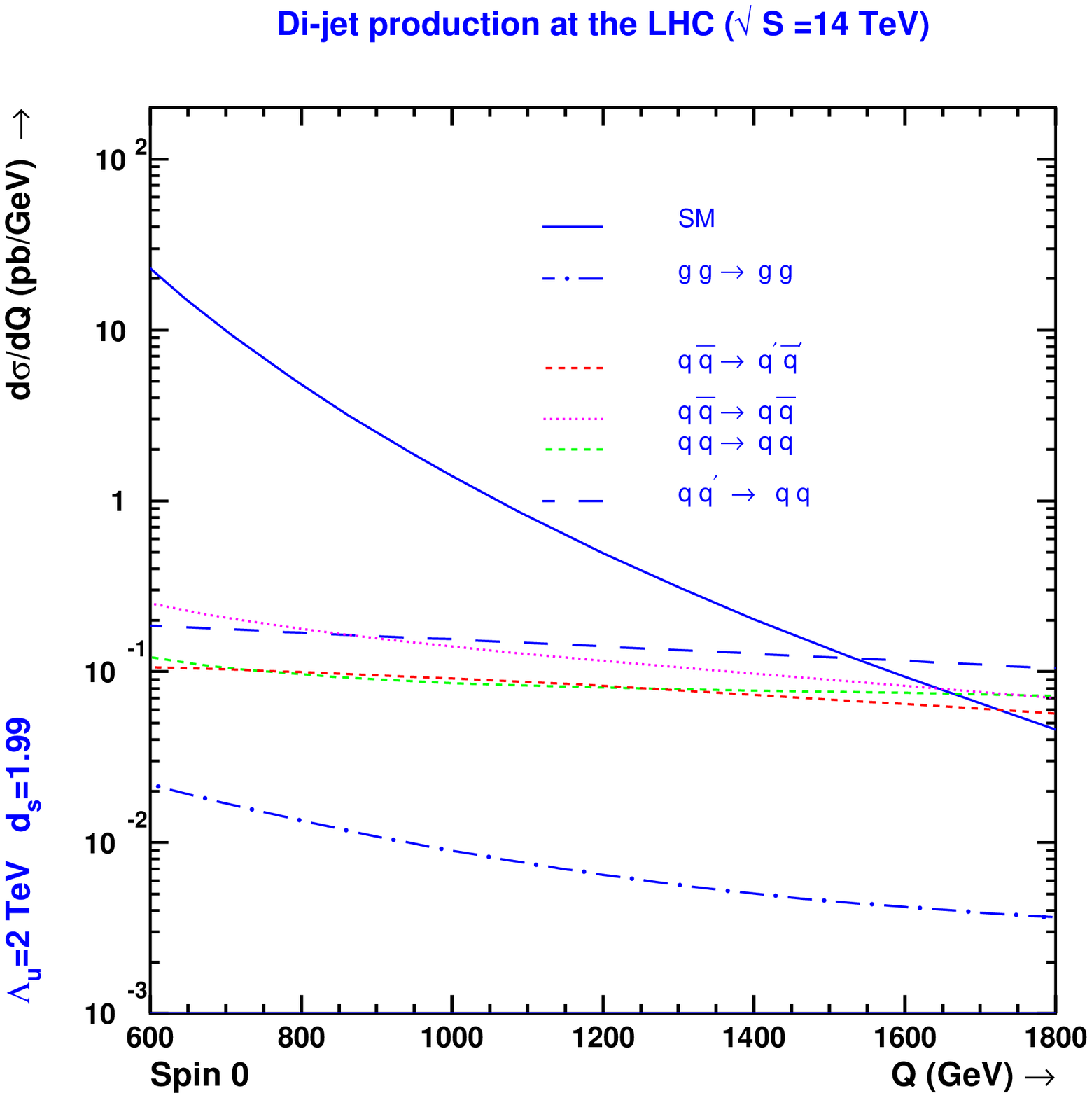,width=9cm,height=9cm,angle=0}
\epsfig{file=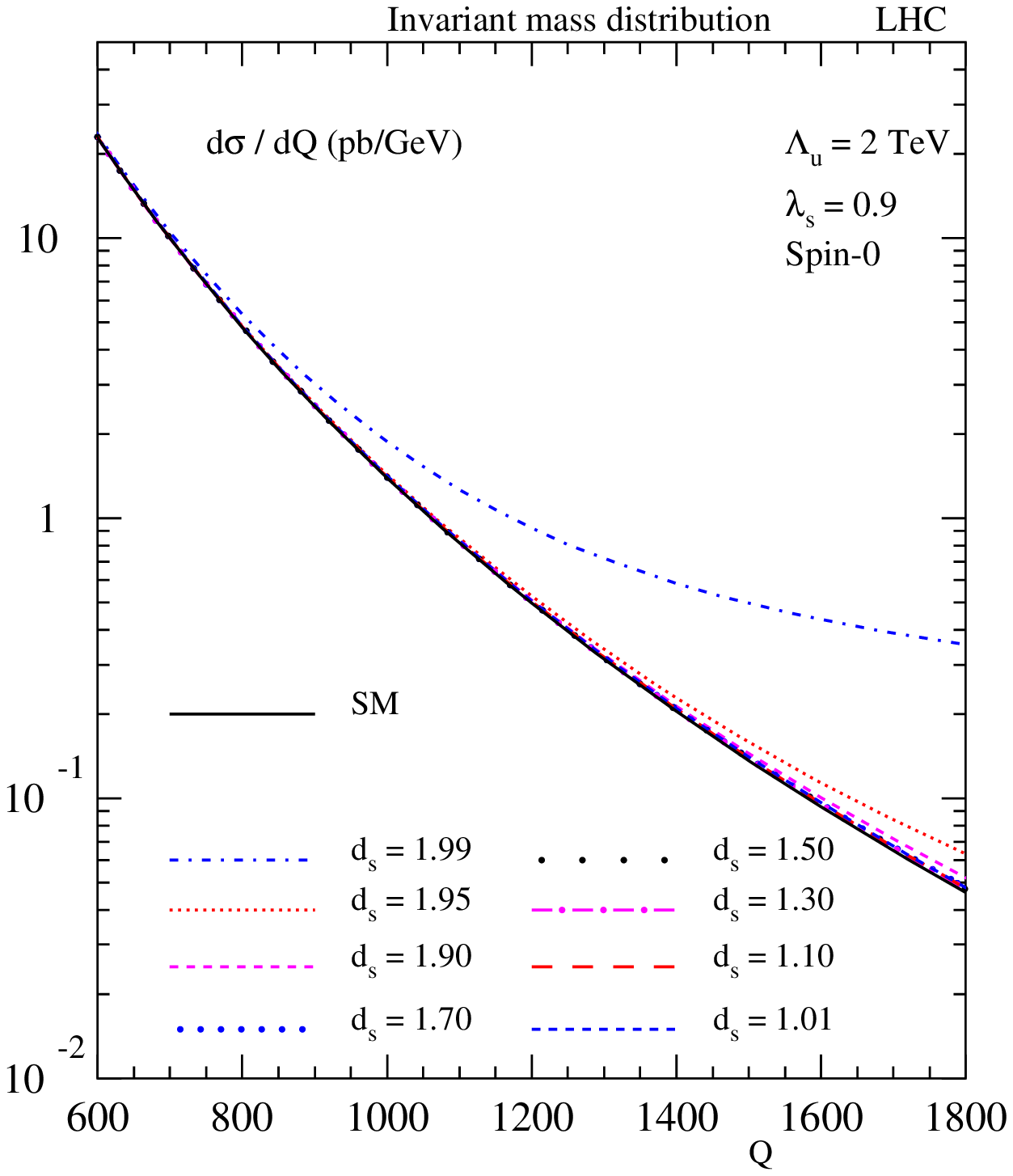,width=7.5cm,height=8.5cm,angle=0}
}
\caption{
 $ d \sigma/dQ $ for di-jet production with spin-0 unparticles. The
couplings are taken to be $ \lambda_{s1,s2}= 0.9 $ 
Left panel: Subprocess contribution.  
Right panel: Variation with scaling dimension.
}
\label{scalar}
\end{figure}

In the left panel of Fig.\ \ref{scalar}, we have plotted various 
subprocess contributions to the di-jet invariant mass distribution 
for spin-0 case (Eq.\ (\ref{Lint})) as a function of $Q$ between
$600~$GeV and $1800~$GeV.  Interference of unparticle contribution 
with SM is added to those coming from pure unparticles in plotting 
these curves.  Solid line gives the prediction in the SM.  We note 
that $ gg \rightarrow gg $ contribution resulting from the first 
term of Eq.\ (\ref{Lint}) is numerically small compared to those 
coming from the second term.  All the 
quark (and anti-quark) initiated subprocesses contribute
almost equally for most of the $Q$ range.  In the right panel, we have 
shown the sensitivity of the scaling dimension $d_s$ to $ d\sigma/dQ$ distribution as
a function of $Q$.
Various curves here correspond to different values of $d_s~$ in the range 1.0-2.0.
It is clear from the right panel that the 
unparticle effects in the fermion initiated processes can be
visible above $Q=800$~GeV for scaling dimension closer to 1.99.  As we deviate
from $d_s=1.99$ to lower values, the effects get washed away completely 
for a wide range of $Q$.

\begin{figure}[htb]
\vspace{1mm}
\centerline{
\epsfig{file=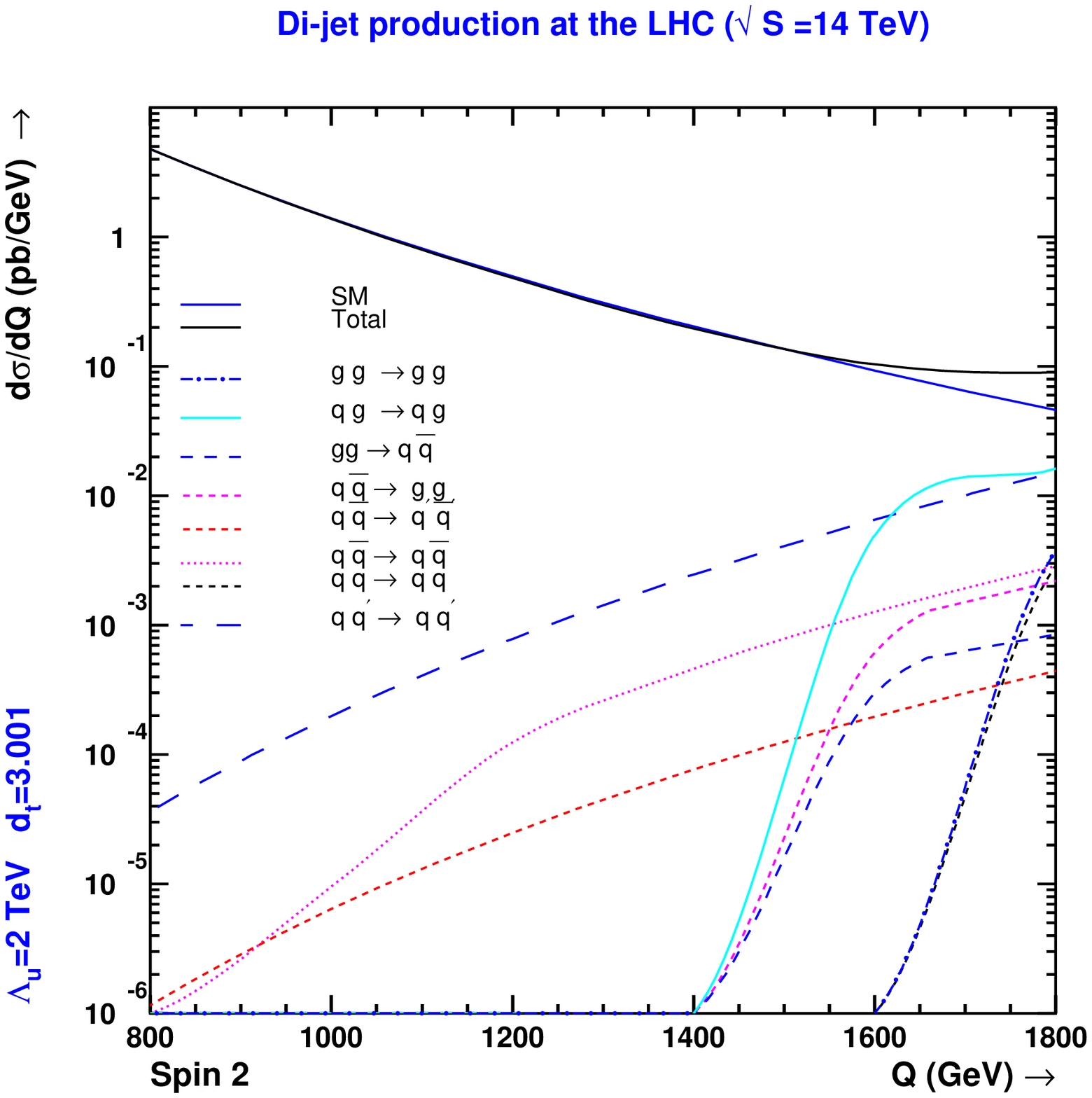,width=9cm,height=9cm,angle=0}
\epsfig{file=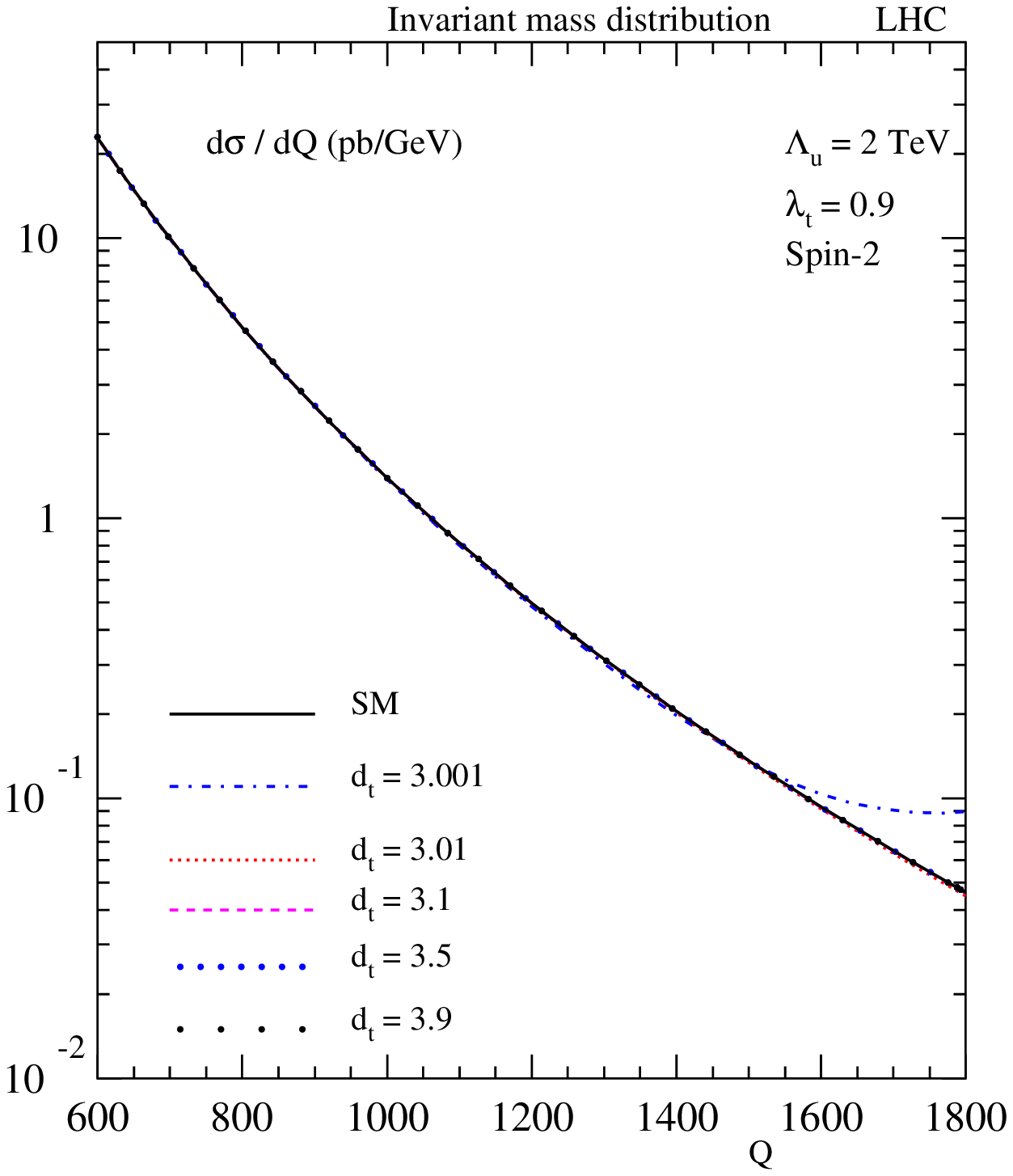,width=7.5cm,height=8.5cm,angle=0}
}
\caption{Spin-2 subprocess contributions
to the di-jet invariant mass 
distribution for $\Lambda_u$=2TeV and $d_t=3.001$.}
\label{tsub}
\end{figure}

In Fig.\ \ref{tsub} we have plotted the invariant mass distribution 
resulting from the spin-2 unparticles (see Eq.\ (\ref{Lintt})).  Scale 
invariance restricts the scaling dimension to be greater than 3.  In 
the left panel we have used $d_t =3.001$ for giving various subprocess
contributions. In the right panel of fig (\ref{tsub}) we show the i
variation of the signal with varying $d_t$ over a wide range 3.0-4.0.
We find that the spin-2 unparticles do not give significant enhancement 
over the SM di-jet cross-section.
The tensor unparticle contribution is smaller than that of
the scalar unparticle due to the additional $\Lambda_u$ suppression
(see Eq.\ (\ref{Lintt})) and large $d_t$ value. 
This is in contrast to the di-photon production, for example,
which gets significant enhancements from spin-2 unparticles as well.

\begin{figure}[htb]
\centerline{
\epsfig{file=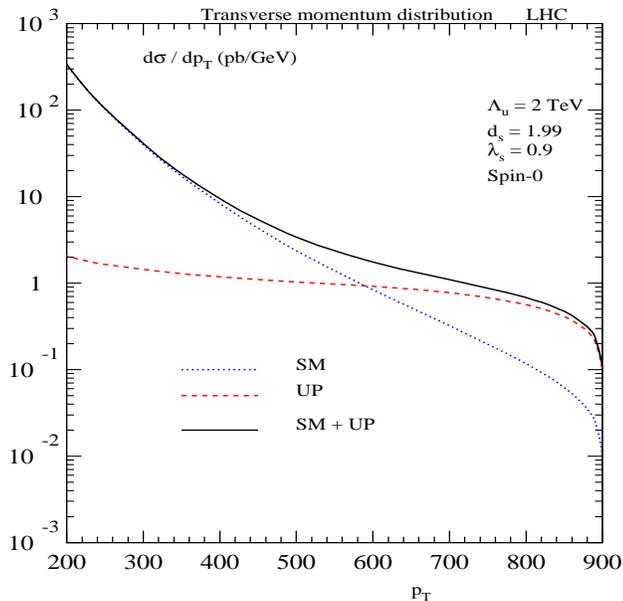,width=8.5cm,height=8.5cm,angle=0}
}
\caption{Scalar unparticle contribution to 
transverse momentum distributions of the jet in the di-jet production for 
$\Lambda_{u}=2$ TeV for $d_s=1.99$.} 
\label{pts0}
\end{figure}

\begin{figure}[htb]
\centerline{
\epsfig{file=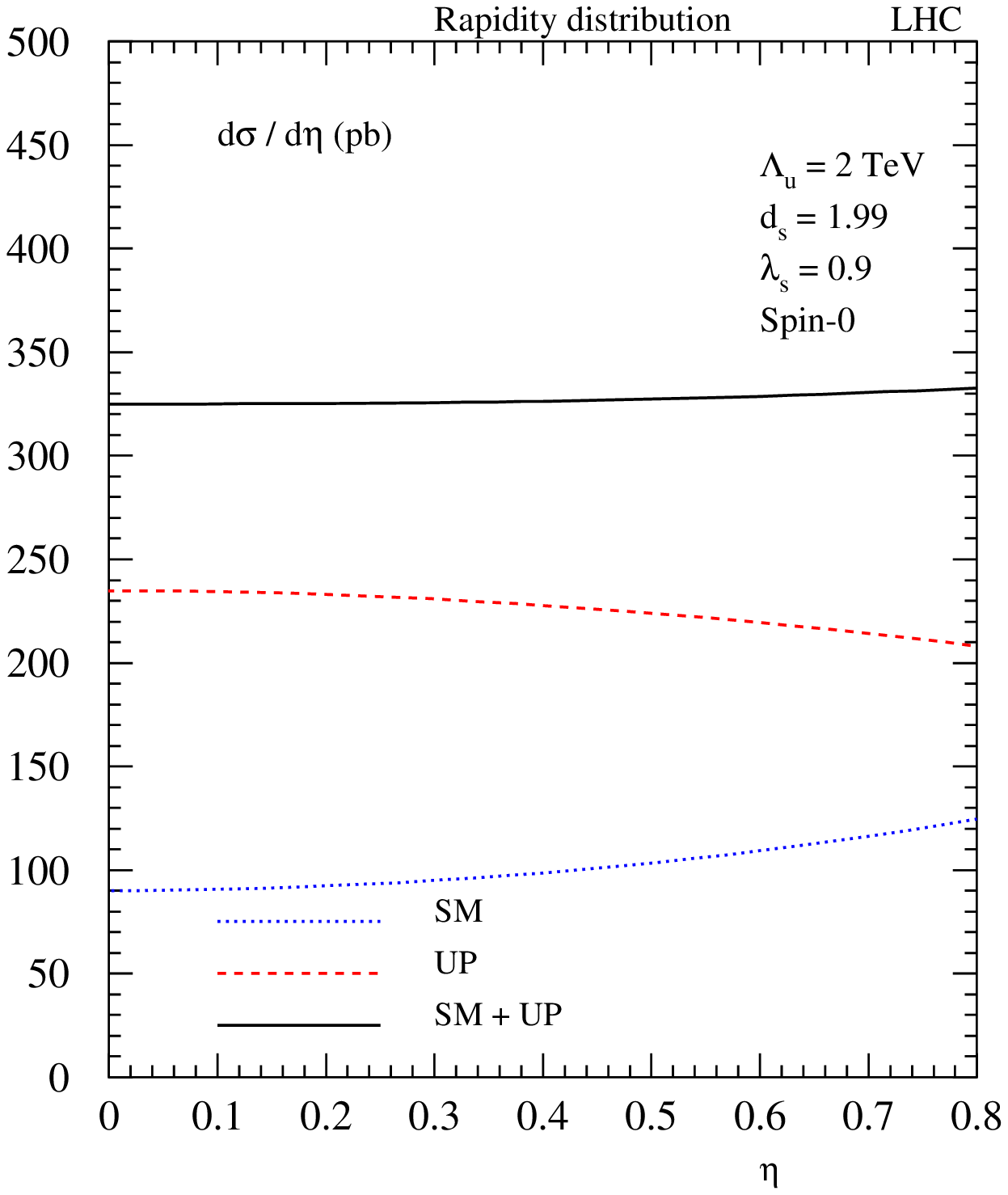,width=8.5cm,height=8.5cm,angle=0}
\epsfig{file=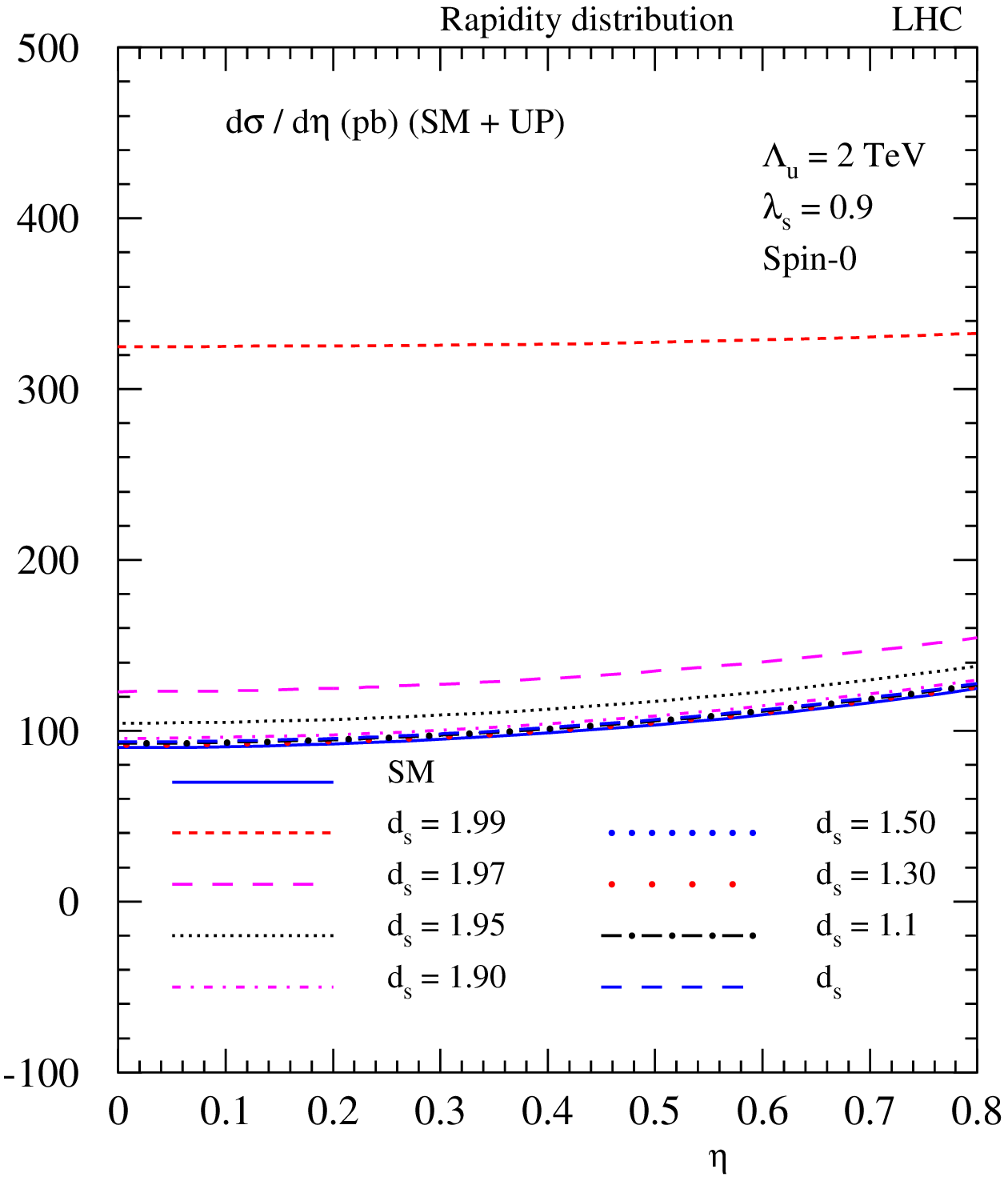,width=8.5cm,height=8.5cm,angle=0}
}
\caption{Scalar unparticle contribution to 
rapidity distributions of the jet in the di-jet production for 
$\Lambda_{u}=2$ TeV. Left panel: Individual contributions
for $d_s=1.99$. Right panel: Variation of signal with $d_s$}
\label{ys0}
\end{figure}

In Fig.\ \ref{pts0}, we present the scalar unparticle contribution
in the transverse momentum distributions of the jets.  
Because of the rapidity cut $|\eta| < 1$ on the jets, the events will be
transverse in nature.
In the limit where the momentum of the jet is in the transverse direction, 
it is easy to see that $p_T=Q/2$, where $Q$ is the invariant mass 
of the di-jet. As $Q$ is required to be less than $\Lambda_{u}$ 
in the unparticle sector, we choose $Q^{max}=0.9~\Lambda_{u}$ 
which translates into $p_T^{max}=Q^{max}/2$.   As $p_T$ is directly related
to $Q$, 
the unparticle
contribution is expected to be visible in the high $p_T$ region as can be
seen from the figure (left panel).  The steep fall in the distribution 
close to 900 GeV is due to the limit on $p_T~(<p_T^{max})$. 

In Fig.\ \ref{ys0} we present rapidity distributions for scalar unparticles.
In evaluating this distribution for non-identical final state partons, we 
have added to the matrix element square $|M(t,u)|^2$ the $t\leftrightarrow 
u$ piece $|M(u,t)|^2$ 
as experimentally it is difficult to distinguish between different partonic 
jets.  We have integrated over $Q$ in the region $1200 < Q < Q^{max}$ where 
the unparticle contribution is dominant over the SM background and present
the rapidity distribution in the range $0 < \eta < 0.8$ (Fig.\ \ref{ys0}).

\begin{figure}[htb]
\vspace{1mm}
\centerline{
\epsfig{file=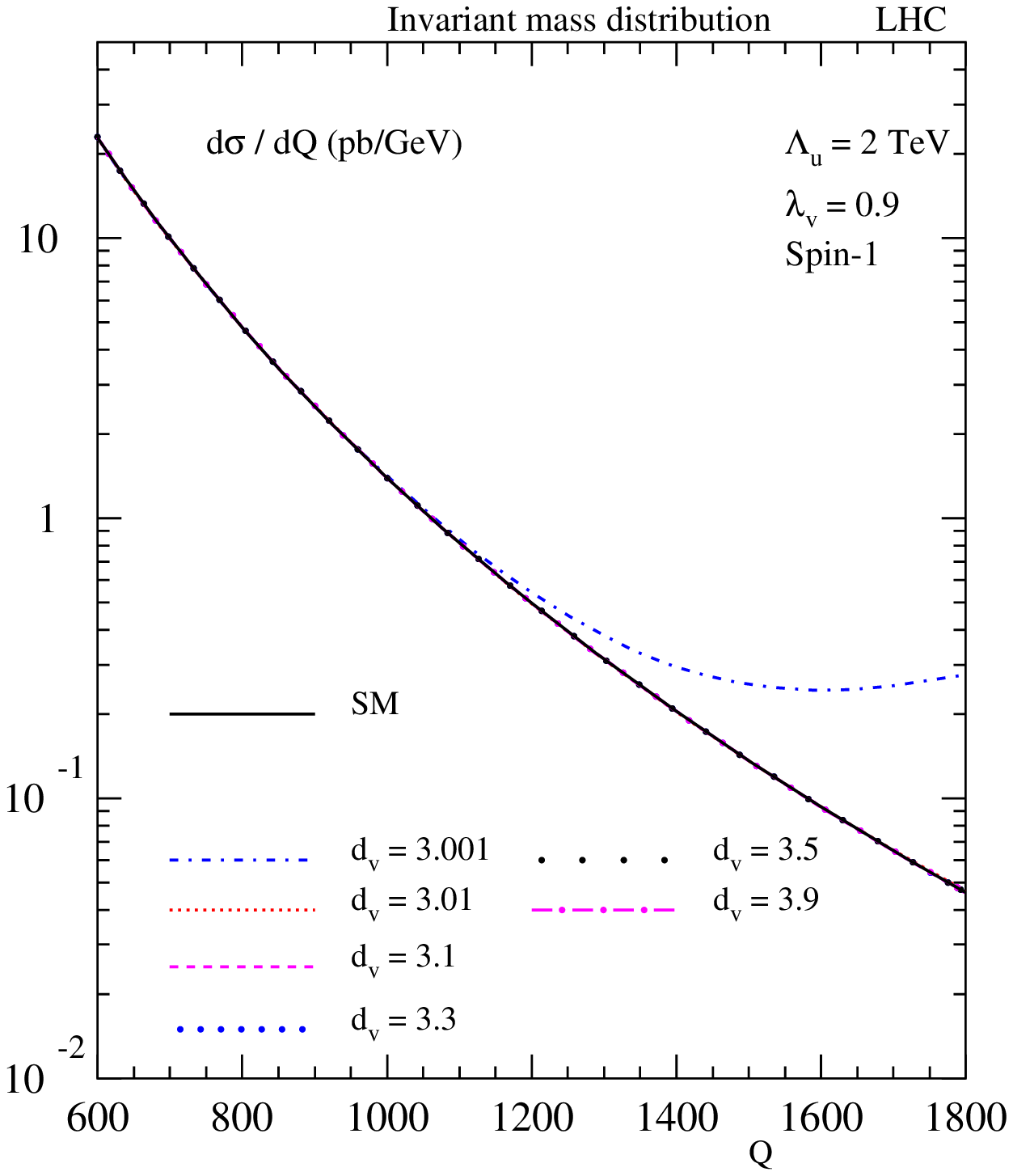 ,width=7.5cm,height=8.5cm,angle=0}
\epsfig{file=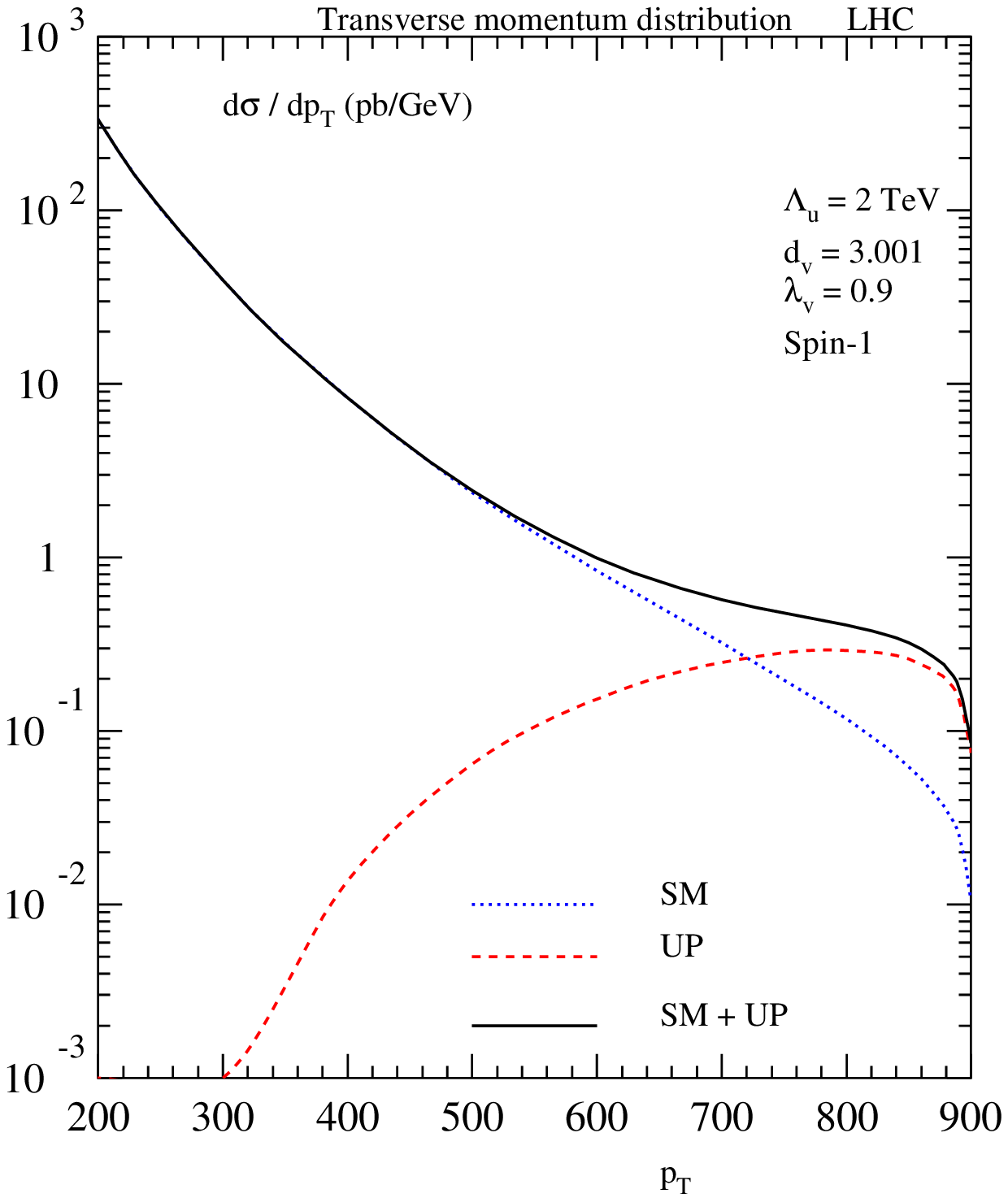,width=7.5cm,height=8.5cm,angle=0}
}
\caption{
Invariant di-jet mass
distribution and $p_T$ distribution for spin-1 unparticles for
$\Lambda_u$ =2TeV. Left Panel: Variation of the signal with 
scaling dimension $d_v$. Right panel: $p_T$ distribution
for $d_v=3.001$}
\label{vQpt}

\vspace{1mm}
\centerline{
\epsfig{file=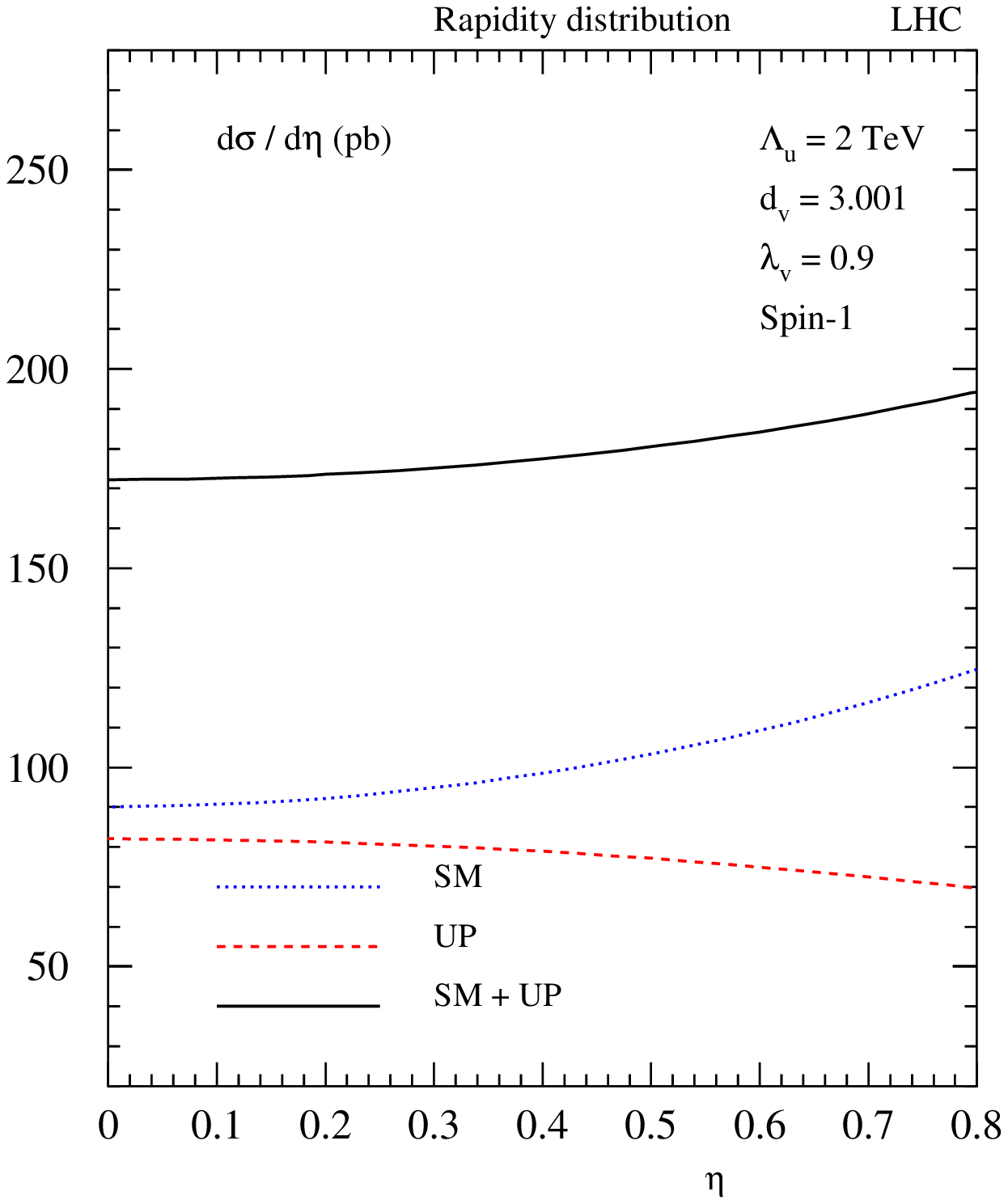,width=7.5cm,height=8.5cm,angle=0}
}
\caption{ Rapidity distribution for
spin-1 unparticles for $\Lambda_u$ =2TeV and   $d_v=3.001$
}
\label{vy}
\end{figure}

Finally we present the results for spin-1 unparticles.  In Fig.\ \ref{vQpt} 
we present invariant mass and $p_T$ distributions for spin-1 unparticles.  
In the left panel we show the variation of invariant mass distribution with 
the scaling dimension $d_v$ and in the right panel for $d_v=3.001$ we have 
plotted the $p_T$ distributions. In Fig.\ \ref{vy} the rapidity distribution 
is plotted in the range $0<\eta<0.8$ for $d_v=3.001$.  We note that the 
signal is not significantly different from SM unless we are very close to 
$d_v=3$.

To summarise, in this article, we have studied the effects of scalar, 
spin-1 and spin-2 unparticles to the di-jet production at the LHC.  
We have considered two SM scalar operators constructed out of quark 
(anti-quark) and gluon field operators that couple to the scalar 
unparticles.  The SM operator which couples to the vector unparticle 
operator is constructed out of the quark (anti-quark) fields.  The 
spin-2 tensor unparticles couple to a second rank SM operator which 
we take it to be SM energy-momentum tensor.  We have computed all 
the parton level cross sections and implemented in a Monte Carlo code 
that can easily incorporate all the experimental cuts.  We have 
presented various subprocess contributions coming from scalar
and tensor unparticles along with the SM contributions.  This includes
the interference of unparticle effects with the SM.  We find that
the scalar unparticle effects are larger compared to those of spin-1 
and spin-2 unparticles.  Moreover, for a wide range of $Q$ and scaling 
dimensions, the di-jet cross section is insensitive to spin-1 and spin-2 
unparticles.  However,  the scalar unparticles can give significant 
enhancements in the invariant mass, rapidity and $p_T$ distributions 
over the SM predictions.
\\

\noindent
{\bf Acknowledgments:}
The work of NA and MCK is supported by CSIR Senior Research Fellowship, New Delhi and
NA would also like to thank Dr. V.K Tiwari for his support and encouragement.
The work of NA, VR
and AT has been partially supported by the  RECAPP, the Department of
Atomic Energy, Govt.  of India.  NA, AT and VR would also like to thank
the cluster computing facility at Harish-Chandra Research Institute.

\begin{table}[h]
\begin{center}
\begin{tabular}{|l|l|}
\hline
{} & {} \\ [2ex]
subprocess  & $\overline \sum |M|^2$\\
{} & {} \\ [2ex]
\hline
{} & {} \\ [2ex]
$q \qp \rarrow q \qp $ &
${\displaystyle \frac{f_N}{2}  ~ \frac{ s^2 + u^2 }{t^2}}$ \\ [4ex]
$q q \rarrow  q q$  &
 ${\displaystyle \frac{f_N}{2} ~ \left( \frac{u^2 + s^2}{t^2} + \frac{t^2 + s^2}{u^2} \right)
             - \frac{f_N}{N} ~ \frac{s^2}{ut}}$ \\ [4ex]
$q \qb \rarrow q \qb$  &
               ${\displaystyle \frac{f_N}{2} ~ \left( \frac{u^2 + t^2}{s^2} + \frac{u^2 + s^2}{t^2
} \right)
             - \frac{f_N}{N} ~ \frac{u^2}{st}}$ \\ [4ex]
$q \qb \rarrow \qp \qbp$  &
${\displaystyle \frac{f_N}{2} ~ \frac{u^2 + t^2}{s^2}}$ \\ [4ex]
$q \qb \rarrow g g$   &
${\displaystyle \frac{N f_N}{2} ~ \frac{ (u^2 + t^2)^2 }{u t s^2}  - \frac{f_N}{2 N}~  \frac{u^2 +
 t^2}{u t}}$ \\ [4ex]
$gg \rarrow q \qb$  &
${\displaystyle \frac{1}{2 N}~  \frac{ (u^2 + t^2)^2 }{u t s^2}  -\frac{1}{N^3 f_N} ~  \frac{u^2 +
 t^2}{s^2}}$ \\ [4ex]
$qg \rarrow qg$   &
${\displaystyle \frac{ (u^2 + s^2) }{t^2}  -\frac{f_N}{2}~   \frac{u^2 + s^2}{us}}$ \\ [4ex]
$gg \rarrow gg$  &
${\displaystyle \frac{4}{f_N}~  \frac{ \left( s^2 + su + u^2 \right)^3 }{s^2 u^2 t^2} } $\\[4ex]
\hline
\end{tabular}
\end{center}
\caption{Matrix elements for the Standard Model.  $s = ( p_1 + p_2 )^2,t= (p_1 - p_3)^2 ,
u =(p_1 - p_4)^2$ and $f_N = (N^2-1)/N^2$}
\label{smm}
\end{table}
\begin{table}[h]
\begin{center}
\begin{tabular}{|l|l|l|}
\hline
{} & {}& \\ [2ex]
subprocess  & $\overline \sum |{\cal M}_{u}|^2$   &   $2~ {\cal R}e \Big({\cal M}_{u}
{\cal M}^{*}_{SM}\Big) $  \\
{} & {}& \\ [2ex]
\hline
{} & {}& \\ [2ex]
$q \qp \rarrow q \qp$
        & ${\displaystyle {\cal D}_t^2 ~t^2}$
        & 0 \\ [4ex]
$q q \rarrow  q q$
        & ${\displaystyle \frac{1}{N}{\cal D}_t {\cal D}_u ~ t u
          +  {\cal D}_t^2 ~ t^2
            + {\cal D}_u^2~ u^2 }$
& ${\displaystyle f_N~\Big( {\cal D}_u ~  \frac{u^2}{t}
          +  {\cal D}_t~ \frac{t^2}{u}\Big)} $\\ [4ex]
$q \qb \rarrow q \qb$
       &  $\displaystyle{\frac {1}{ N} ~ {\cal D}_t Re({\cal D}_s) ~ s t
           +  {\cal D}_t^2 ~ t^2
           + |{\cal D}_s|^2 ~s^2} $
       & $\displaystyle{f_N ~\Big(  Re({\cal D}_s) ~ \frac{s^2}{t}
           +  {\cal D}_t ~ \frac{t^2}{s}\Big)} $\\ [4ex]
$q \qb \rarrow \qp \qbp$
& ${\displaystyle |{\cal D}_s|^2 ~ s^2}$
& 0 \\[2ex]
\hline
\end{tabular}
\caption{Matrix elements for scalar unparticle
 in fermion initiated processes $f_N = (N^2-1)/N^2$ }
\label{smatrix}
\end{center}
\end{table}
\begin{table}[h]
\begin{center}
\begin{tabular}{|l|l|l|}
\hline
{} & {}& \\ [2ex]
subprocess  & $\overline \sum |{\cal M}_{u}|^2$   &   $2~ {\cal R}e \Big({\cal M
}_{u}
{\cal M}^{*}_{SM}\Big) $  \\
{} & {}& \\ [2ex]
\hline
{} & {}& \\ [2ex]
$q \qp \rarrow q \qp$
        & ${\displaystyle 2{\cal D}_t^2
          ~\left(2 u^2+ 2 u t + t^2 \right) }$             & 0
\\ [4ex]
$q q \rarrow  q q$
        & ${\displaystyle 2 \Big(
              \frac{2}{N}{\cal D}_t {\cal D}_u ~ \left( u^2 +2 u t +  t^2 \right
)
             + {\cal D}_t^2 ~ \left( 2 u^2 + 2 t u + t^2 \right) }$
                                                           & ${\displaystyle -2
g_s^2 f_N~ s^2  ~\Big(
                                                              \frac{{\cal D}_u}{
t}
                                                           +  \frac{{\cal D}_t}{
u} \Big)} $\\ [2ex]
     & \hspace{0.8cm}${\displaystyle + {\cal D}_u^2~  \left( u^2 +2 t u + 2 t^2
\right) \Big)}    $ &
\\ [4ex]
$q \qb \rarrow q \qb$
       &  $\displaystyle { 2 \Big(
              \frac {2}{ N} ~ {\cal D}_t Re({\cal D}_s) ~  u^2
           +  {\cal D}_t^2 ~ \left( 2 u^2 + 2 t u + t^2 \right) }$
                                                             & $\displaystyle{ -
2 g_s^2 f_N~  u^2 ~\Big(
                                                                \frac{Re({\cal D
}_s)}{t}
                                                             +  \frac{{\cal D}_t
}{s}                  \Big)} $\\ [2ex]
    & \hspace{0.8cm}${\displaystyle + |{\cal D}_s|^2 ~ \left( u^2 +t^2 \right) \
Big) } $ &
\\ [4ex]
$q \qb \rarrow \qp \qbp$
& ${\displaystyle 2 |{\cal D}_s|^2 ~ \left( u^2+ t^2 \right) }$
& 0 \\[2ex]
\hline
\end{tabular}
\caption{Matrix elements for spin-1 unparticle
 in fermion initiated processes $f_N = (N^2-1)/N^2$ }
\label{somatrix}
\end{center}
\end{table}
\vspace{10pt}
\begin{table}[hbp]
\begin{center}
\begin{tabular}{|l|l|l|}
\hline
{} & {}& \\ [2ex]
subprocess  & $\overline \sum |{\cal M}_{u}|^2$   &   $2~{\cal R}e \Big({\cal M}_{u} {\cal M}^{*}_{SM}\Big) $  \\
{} & {}& \\ [2ex]
\hline
{} & {}& \\ [2ex]
\qqpZqqp & $ {\displaystyle {1 \over 128} {\cal D}_t^2~\Big( u^4 + s^4 ~ 
                     -6 su \big[ u^2+ s^2 - 3su \big] \Big)}$ &   0
\\[4ex]
$\qqZqq$ & ${\displaystyle  {1 \over 128 N} \Big(N {\cal D}_t^2 
                \big[u^4+s^4 -6 s u~\big(u^2+s^2-3 s u\big)\big]}$                 & ${\displaystyle -\frac{1}{8} {\cal D}_u f_N 
                                                            ~ ( 3u + 4s ) \frac{s^2}{t} 
                                                            - (u \leftrightarrow t)}$
                                                               \\[2ex]
         & \hspace{0.8cm}${\displaystyle + {\cal D}_t {\cal D}_u s^2 \big[4 s^2 
           +9 u t\big]\Big) + t\leftrightarrow u}$         &
\\[4ex]
$\qqbZqqb$ &  ${\displaystyle {1 \over 128 N} \Big(N D_t^2 \big[u^4+s^4
            -6 s u~ \big(u^2+s^2-3 s u\big)\big]}        $ & ${\displaystyle -\frac{1}{8} f_N \Big({\cal R}e({\cal D}_s) 
                                                           ~ ( 3s + 4u ) \frac{u^2}{t}}$ \\[2ex]
           & \hspace{0.8cm} ${\displaystyle +{\cal R}e\left({\cal D}_t {\cal D}_s\right) u^2  
              \big[9 t s+4 u^2\big]\Big)              
              +t \leftrightarrow s}$                   & \hspace{0.8cm}${\displaystyle -{\cal D}_t
                                                           ~( 3t + 4u ) \frac{u^2}{s}\Big)}$
\\[4ex]
\qqbZqpqbp & ${\displaystyle \frac{1}{128} |{\cal D}_s|^2 
             ~ \Big( u^4 +t^4 ~ -6 tu 
              \left[ u^2 + t^2 - 3tu \right] \Big)}$ & 0
\\ [4ex] 
\qqbZgg  & ${\displaystyle \frac{1}{8} N f_N |{\cal D}_s|^2 
          ~  ut  \Big( u^2 + t^2 \Big)}$            & ${\displaystyle -{1 \over 2} f_N  {\cal R}e({\cal D}_s) 
                                                        ~ \Big(u^2 + t^2\Big)} $ 
 \\ [4ex] 
\ggZqqb & ${\displaystyle \frac{1}{8} \frac{1}{N f_N} |{\cal D}_s|^2 
          ~  ut  \Big( u^2 + t^2 \Big)} $           & ${\displaystyle  - {1 \over 2 (N^2-1)}{\cal R}e({\cal D}_s)~   
                                                       \Big(u^2 + t^2\Big)} $ 
\\ [4ex] 
\qgZqg   & ${\displaystyle -\frac{1}{8} {\cal D}_t^2 
          ~ us  \Big( u^2 + s^2 \Big)}$   & ${\displaystyle  {1 \over 2 N}{\cal D}_t 
                                                ~  \Big(u^2 + s^2\Big)} $ 
 \\ [4ex] 
\ggZgg  & ${\displaystyle \frac{2}{8(N^2-1)} 
           ~ \Big( {\cal D}_u {\cal R}e({\cal D}_s) ~ t^4 
           + {\cal D}_t {\cal R}e({\cal D}_s)~ u^4} $      & ${\displaystyle -\frac{1}{N f_N} 
                                                     \Big( {\cal D}_u \frac{ t^4 + s^4 }{st}
                                                       +{\cal D}_t \frac{ u^4 + s^4 }{su}} $\\
       &\hspace{0.8cm}${\displaystyle  + {\cal D}_t {\cal D}_u ~ s^4 \Big) 
         + {1 \over 8} {\cal D}_u^2 
         ~ \left( t^4 + s^4 \right)}$               & ${\displaystyle  \hspace{0.8cm}+{\cal R}e({\cal D}_s) \frac{ t^4 + u^4 }{ut} 
                                                    \Big)}$\\
       &\hspace{0.8cm}${\displaystyle   + {1 \over 8} {\cal D}_t^2 
         ~ \left( u^4 + s^4 \right)
         + {1 \over 8} |{\cal D}_s|^2 
          ~ \left( t^4 + u^4 \right)}$           & 
\\[2ex]
\hline
\end{tabular}
\end{center}
\caption{Matrix elements for spin 2 unparticle}
\label{tt}
\end{table}



\begin{thebibliography}{99}

\bibitem{Hofman:2008ar}
  D.~M.~Hofman and J.~Maldacena,
  JHEP {\bf 0805} (2008) 012
  [arXiv:0803.1467 [hep-th]].

\bibitem{Georgi:2007ek}
  H.~Georgi,
  Phys.\ Rev.\ Lett.\  {\bf 98} (2007) 221601
  [arXiv:hep-ph/0703260].


\bibitem{Banks:1981nn}
  T.~Banks and A.~Zaks,
  Nucl.\ Phys.\  B {\bf 196} (1982) 189.

\bibitem{Nakayama:2007qu}
  Y.~Nakayama,
  Phys.\ Rev.\  D {\bf 76}, 105009 (2007).
%
\bibitem{Mack:1975je}
  G.~Mack,
  Commun.\ Math.\ Phys.\  {\bf 55}, 1 (1977).
%
\bibitem{Grinstein:2008qk}
  B.~Grinstein, K.~A.~Intriligator and I.~Z.~Rothstein,
  Phys.\ Lett.\  B {\bf 662} (2008) 367
  [arXiv:0801.1140 [hep-ph]].
%
\bibitem{Georgi:2007si}
  H.~Georgi,
  Phys.\ Lett.\  B {\bf 650} (2007) 275
  [arXiv:0704.2457 [hep-ph]].

\bibitem{Cheung:2007zza}
  K.~Cheung, W.~Y.~Keung and T.~C.~Yuan,
  Phys.\ Rev.\ Lett.\  {\bf 99} (2007) 051803
  [arXiv:0704.2588 [hep-ph]].

\bibitem{Cheung:2008xu}
  K.~Cheung, W.~Y.~Keung and T.~C.~Yuan,
  AIP Conf.\ Proc.\  {\bf 1078} (2009) 156
  [arXiv:0809.0995 [hep-ph]].
  A.~Rajaraman,
  AIP Conf.\ Proc.\  {\bf 1078}, 63 (2009)
  [arXiv:0809.5092 [hep-ph]], and references therein.
%
\bibitem{Alan:2007rg}
  A.~T.~Alan,
  arXiv:0711.3272 [hep-ph].
\bibitem{Mathews:2007hr}
  P.~Mathews and V.~Ravindran,
  Phys.\ Lett.\  B {\bf 657} (2007) 198
  [arXiv:0705.4599 [hep-ph]].

\bibitem{Kumar:2007af}
  M.~C.~Kumar, P.~Mathews, V.~Ravindran and A.~Tripathi,
  Phys.\ Rev.\  D {\bf 77} (2008) 055013
  [arXiv:0709.2478 [hep-ph]].

\bibitem{Kumar:2008dn}
  M.~C.~Kumar, P.~Mathews, V.~Ravindran and A.~Tripathi,
  arXiv:0804.4054 [hep-ph].

\bibitem{Randall:2008rw}
  L.~Randall and D.~Tucker-Smith,
  Phys.\ Rev.\ Lett.\  {\bf 101} (2008) 221803
  [arXiv:0806.1049 [hep-ph]].

\bibitem{Anchordoqui:2008di}
  L.~A.~Anchordoqui, H.~Goldberg, D.~Lust, S.~Nawata, S.~Stieberger and T.~R.~Taylor,
  Phys.\ Rev.\ Lett.\  {\bf 101} (2008) 241803
  [arXiv:0808.0497 [hep-ph]].

\bibitem{Mathews:1999iw}
  P.~Mathews, S.~Raychaudhuri and K.~Sridhar,
  JHEP {\bf 0007}, 008 (2000)
  [arXiv:hep-ph/9904232].

\bibitem{Atwood:1999qd}
  D.~Atwood, S.~Bar-Shalom and A.~Soni,
  Phys.\ Rev.\  D {\bf 62} (2000) 056008
  [arXiv:hep-ph/9911231].

\bibitem{Ghosh:1999ex}
  D.~K.~Ghosh, P.~Mathews, P.~Poulose and K.~Sridhar,
  JHEP {\bf 9911} (1999) 004
  [arXiv:hep-ph/9909567].
%
%
\bibitem{Combridge:1977dm}
  B.~L.~Combridge, J.~Kripfganz and J.~Ranft,
  Phys.\ Lett.\  B {\bf 70} (1977) 234.

\bibitem{Bhatti:2008hz}
  A.~Bhatti {\it et al.},
  J.\ Phys.\ G {\bf 36} (2009) 015004
  [arXiv:0807.4961 [hep-ex]].

\end{thebibliography}
\end{document}